\documentclass[article,aps,pra,twocolumn,longbibliography,superscriptaddress]{revtex4-1}  
\newtheorem{thm}{Theorem}

\usepackage{amsmath}    
\usepackage{amsfonts} 
\usepackage{amssymb}
\usepackage{graphicx}   
\usepackage{algorithm}
\usepackage[colorlinks,breaklinks=true]{hyperref}
\usepackage{algpseudocode}

\newcommand{\ket}[1]{\left|#1\right\rangle}

\newcommand{\1}{\ket{1}}

\newcommand{\ord}{\operatorname{ord}}

\begin{document}

\title{A quantum primality test with order finding}
\author{Alvaro Donis-Vela}
\affiliation{Universidad de Valladolid, G-FOR: Research Group on Photonics, Quantum Information and Radiation and Scattering of Electromagnetic Waves.}
\author{Juan Carlos Garcia-Escartin}
 \email{juagar@tel.uva.es}  
\affiliation{Universidad de Valladolid, G-FOR: Research Group on Photonics, Quantum Information and Radiation and Scattering of Electromagnetic Waves.}
\affiliation{Universidad de Valladolid, Dpto. Teor\'ia de la Se\~{n}al e Ing. Telem\'atica, Paseo Bel\'en n$^o$ 15, 47011 Valladolid, Spain}
\date{\today}

\begin{abstract} 
Determining whether a given integer is prime or composite is a basic task in number theory. We present a primality test based on quantum order finding and the converse of Fermat's theorem. For an integer $N$, the test tries to find an element of the multiplicative group of integers modulo $N$ with order $N-1$. If one is found, the number is known to be prime. During the test, we can also show most of the times $N$ is composite  with certainty (and a witness) or, after $\log\log  N$ unsuccessful attempts to find an element of order $N-1$, declare it composite with high probability. The algorithm requires $O((\log n)^2 n^3)$ operations for a number $N$ with $n$ bits, which can be reduced to $O(\log\log n (\log n)^3 n^2)$ operations in the asymptotic limit if we use fast multiplication.
\end{abstract}
\maketitle
Prime numbers are the fundamental entity in number theory and play a key role in many of its applications such as cryptography. Primality tests are algorithms that determine whether a given integer $N$ is prime or not. A na\"ive but inefficient solution is trying all the numbers up to $\sqrt{N}$ looking for a factor, which would prove $N$  is prime if no factor is found and show it is composite if we have one. There are more efficient ways to test for primality based on different results from number theory. We are going to use basic theorems which can be found, together with their proofs, in elementary number theory books \cite{MVS96,Bur05,CP05}.

Some definitions are useful before we proceed. Let $\mathbb{Z}_N$ be the ring of integers modulo $N$ and $(a,N)$ the greatest common divisor of $a$ and $N$. We call $\mathbb{Z}_N^*$ to the \emph{multiplicative group of integers modulo N} defined as $\mathbb{Z}_N^*=\{a \in \mathbb{Z}_N : (a,N)=1\}$. The elements of $\mathbb{Z}_N^*$ are the integers from $1$ to $N-1$ which are coprime to $N$. These integers form a group under multiplication. 

The order of a finite group $G$, $|G|$, is the number of elements of that group (its cardinality). The order of $\mathbb{Z}_N^*$ is given by Euler's totient function $\varphi (N)$ which gives how many integers $1 \leq a < N$ are coprime to $N$.

The multiplicative order of an element $a \in \mathbb{Z}_N^*$, $ord(a)$, is the smallest positive integer $r$ such that $a^r\equiv 1 \mod N$. 

With these concepts and known theorems from number theory, we can give different tests to check if an integer is prime or not. A simple test is given by Fermat's theorem:
\begin{thm}[Fermat]\label{FermatTh}
If a positive integer $N$ is prime, then 
$a^{N-1}\equiv 1 \mod N$
for any positive integer $a$ such that $(a,N)=1$.
\end{thm}
This is a special case of Euler's theorem:
\begin{thm}[Euler]\label{EulerTh}
Let $N$ be a positive integer, then 
$a^{\varphi (N)}\equiv 1 \mod N$
for any positive integer $a$ such that $(a,N)=1$.
\end{thm}
For a prime $N$, $\varphi(N)=N-1$ and we recover Fermat's theorem. 

If we can find an integer $a$ for which $a^{N-1}\not\equiv 1 \mod N$, we have proof $N$ is composite and we call $a$ a Fermat witness for compositeness. Given $a$, anyone can quickly check $N$ is not prime. This gives a simple test for primality. We pick a random $a$ from $1$ to $N$, verify $(a,N)=1$ (otherwise we know $N$ is composite) and then check for Fermat's condition. After testing a few different elements, we can declare it prime with high probability.

While this test is simple, there are certain numbers, called Carmichael numbers \cite{Car12,AGP94}, which obey Fermat's condition for every possible $a$. Any other integer will fail Fermat's test at least half of the times. To see this, we can use Lagrange's theorem:
\begin{thm}[Lagrange]\label{LagrangeTh}
Let $|G|$ be the number of elements of a finite group $G$, then any subgroup $S$ of $G$ must have a number of elements $|S|$ which is a divisor of the size of the group.
\end{thm}
There is a direct application of Lagrange's theorem to Fermat's test. Fermat liars are the integers $a$ such that $a^{N-1}\equiv 1 \mod N$ for a composite $N$. The liars form a subgroup of $\mathbb{Z}_N^*$. If $a_1^{N-1}\equiv 1 \mod N$ and $a_2^{N-1}\equiv 1 \mod N$ then $a_3=a_1a_2$ is also a liar, but multiplication outside the subgroup breaks closure. The remaining subgroup properties, like the existence of an inverse, an identity, associativity and commutativity, come immediately from the properties of $\mathbb{Z}_N^*$ and the multiplication property we just described. 

From Lagrange's theorem, we see the subgroup $\mathbb{L}$ of Fermat liars must have a size which is a divisor of the size of $\mathbb{Z}_N^*$. The number of possible liars divides $\varphi(N)$ and, if there is at least one Fermat witness which shows $N$ is composite, there must be at least $\varphi(N)/2$ witnesses. $|\mathbb{L}|$ must have a number of elements $\varphi(N)/d$, where $d | \varphi(N)$ ($d$ is divisor of $\varphi(N)$). For $|\mathbb{L}|\neq \varphi(N)$, $d>1$ and $|\mathbb{L}|$ is always equal to or smaller than $\varphi(N)/2$. 

There are also more refined probabilistic tests similar to Fermat's which are based on more sophisticated properties. Most take advantage of Lagrange's theorem to show there is at least one witness and, therefore, the subgroup of liars has a cardinality of, at most, $|\mathbb{Z}_N^*|/2$. By choosing a random $a$ for the test, after a few attempts with different elements, we either find a witness of compositeness or we can be satisfied that there is an exponentially small probability the number is not a prime. The two most important such methods are the Solovay-Strassen test \cite{SS77} and the Miller-Rabin test \cite{Mil75,Rab80} with liar subgroups at most a half and a quarter of the total size respectively. A good description of these and other probabilistic primality tests can be found in Dixon's review \cite{Dix84}.

Notably, there is also a deterministic algorithm for primality testing. The AKS (Agrawal-Kayal-Saxena) primality test \cite{AKS04} is based on a generalization of Fermat's theorem which states that:
\begin{thm}\label{AKS}
A positive integer $N$ is prime if and only if 
\begin{equation}
(x+a)^N\equiv x^N+a \mod N
\end{equation}
for one $a$ such that $(a,N)=1$. 
\end{thm}
The AKS test is deterministic and requires a number of operations essentially of the order of the sixth power of the number of bits of $N$ \cite{LP03}. 

In this paper, we provide a different quantum primality test based on the \emph{converse of Fermat's theorem} \cite{Luc78}
\begin{thm}[Lucas]\label{LucasTh}
If 
\begin{equation}
a^{N-1}\equiv 1 \mod N
\end{equation}
and 
\begin{equation}
a^{x}\not\equiv 1 \mod N
\end{equation}
for any $x< N-1$, then $N$ is prime.
\end{thm}
Apart from Lucas theorem, we make use of a couple of additional results (see chapter 8 of Burton's book \cite{Bur05}):
\begin{thm}\label{OrderTh}
The elements $a \in \mathbb{Z}_N^*$ have an order $\ord(a)| \varphi(N)$ $($the order is always a divisor of $|\mathbb{Z}_N^*|$$)$.
\end{thm}

\begin{thm}\label{OrderPrimesTh}
For a prime $p$ and an integer $d| p-1 $, $\mathbb{Z}_p^*$ has exactly $\varphi (d)$ elements $a$ of order $\ord(a)=d$. 
\end{thm}

We can now restate Lucas theorem as
\begin{thm}[Lucas]\label{LucasAltTh}
If $a\in \mathbb{Z}_N^*$ has order $\ord(a)=N-1$, then $N$ is prime.
\end{thm}
This formulation is contained in Theorem $\ref{OrderTh}$. The order of any element must divide $\varphi(N)$ which is only equal to $N-1$ for prime numbers. The only way we can have order $N-1$ is if $N$ is prime. Additionally, from Theorem $\ref{OrderPrimesTh}$, we see there must be exactly $\varphi(N-1)$ integers with this property. If we can find such an integer, we have a way to prove primality. 

There is no known classical algorithm that can determine the order of an integer efficiently. In order to apply Lucas theorem to primality certification on a classical computer, we need alternative methods. A solution is the Lucas-Lehmer test \cite{Leh27} based on:
\begin{thm}[Lucas-Lehmer]\label{LucasLehmerTh}
If 
\begin{equation}
a^{N-1}\equiv 1 \mod N
\end{equation}
and 
\begin{equation}
a^{\frac{N-1}{p}}\not\equiv 1 \mod N
\end{equation}
for any prime $p | N-1$, then $N$ is prime.
\end{thm}
With this and other refinements, there have been multiple proposals for the efficient implementation of modified Lucas tests on classical computers. In most of them, we require a complete factorization of $N-1$, or, in some, a partial factorization with large factors, both of which might be easier than factoring $N$ (if possible). For instance, these tests are particularly easy to perform on numbers of the form $2^m-1$ \cite{BLS75} which, if prime, are called Mersenne primes and include many of the largest known prime numbers \cite{GIMPS}. The reader can find many of these methods in chapter 4 of Crandall and Pomerance's book \cite{CP05}. 

The tests based on Lucas theorem have the advantage that they allow us to prove primality. With the Solovay-Strassen or the Rabin-Miller test we could only give a witness of compositeness, but there was no efficient way to show $N$ was prime with certainty. 

A \emph{certificate of primality} for $N$ is a collection of data which allows anyone to prove $N$ is prime, ideally with few operations on short certificates with a number of bits of the order of $\log (N)$. For instance, if we have a complete factorization of $N-1$, a list of the factors and an element $a \in \mathbb{Z}_N^*$ with $\ord(a)=N-1$ give a fast way to show $N$ is prime using the Lucas-Lehmer Theorem. In principle, we can always use the AKS test to check if a number is prime and, from a certain point of view, $N$ is itself a valid certificate of primality. However, for large integers, there exist more efficient ways to prove primality if we can factor $N-1$. Pratt certificates were the first examples \cite{Pra75} and there are primality proofs requiring only $O(\log p)$ multiplications modulo $p$ for any prime $p$ \cite{Pom87}.

At this point, it is interesting to turn to quantum computers. Shor's algorithm gives an efficient way to factor composite integers of $n$ bits with a number of expected operations $O((\log n) n^3 )$ operations \cite{Sho97} which can become $O(\log\log n (\log n)^2 n^2 )$ with fast multiplication circuits. The quantum primality test of Chau and Lo \cite{CL97} combines Shor's quantum factoring algorithm with the Lucas-Lehmer test to prove primality in an expected number of operations $O(n^3 \log n \log\log n)$, essentially cubic with the number of bits of $N$. Using quantum factoring has the nice side effect of producing a succint certificate of primality with the results. 

Here, we propose a new quantum primality testing algorithm inspired by Shor's algorithm. By using directly the quantum order finding algorithm behind Shor's factoring and discrete logarithm algorithms, we can reduce the number of quantum operations. Instead of factoring $N-1$, we check the order of different elements in $\mathbb{Z}_N^*$ until we find one with order $N-1$ or a witness that $N$ is composite.

This also contrasts with the quantum primality test of Carlini and Hosoya \cite{CH00}, which applies the concepts of quantum counting and quantum period finding to give an improved version of the Miller-Rabin test.
 
While we reduce the number of quantum operations, we loose the classical certificate of primality of the Chau-Lo test. Instead, we can give a \emph{quantum certificate of primality}. Any of the $\varphi(N-1)$ elements $a\in \mathbb{Z}_N^*$ with $\ord(a)=N-1$ serves to prove $N$ is prime to anyone with a quantum computer, which can find the order of $a$ efficiently.

We deal with numbers $2^{n-1}<N\leq 2^n$ represented with $n$ bits. We consider quantum order finding as a black box which requires $O((\log n)n^3)$ operations and see that, on average, with $\log n$ uses of the black box we can find an element of order $N-1$ and prove $N$ is prime when it is or show it must be composite with high probability, with a proof of compositeness in most cases. 

\begin{algorithm}[H]
\caption{Quantum primality test}
\label{QPrimT}
\begin{algorithmic}[1]
\State Choose at random an integer $1 < a < N$.
\State Compute $(a,N)$:
\If{$(a,N)\neq 1$}
\State Declare $N$ composite; \Return factor $(a,N)$ as a proof.
\ElsIf{$(a,N)=1$}
\State Compute $a^{\frac{N-1}{2}}$:
\If{$a^{\frac{N-1}{2}}\not\equiv \pm 1 \mod N$}
\label{EulerTest}
\State Declare $N$ composite; \Return $a$ as a witness.
\ElsIf{$a^{\frac{N-1}{2}}\equiv 1 \mod N$}
\State Go back to Step 1.
\ElsIf{$a^{\frac{N-1}{2}}\equiv -1 \mod N$}
\State QUANTUM ORDER FINDING. Compute $\ord(a)$:
\If{$\ord(a)=N-1$ }
\State Declare $N$ prime; \Return $a$ as a quantum certificate of primality.
\Else
\State Go back to Step 1.
\EndIf
\EndIf
\EndIf

\end{algorithmic}
\end{algorithm}

The proposed algorithm (Algorithm \ref{QPrimT}) checks the order of random integers in the group until it can prove either primality or compositeness. First we choose an integer at random from $\mathbb{Z}_N^*$, excluding $a=1$ which has a trivial order $1$. We do that by taking an integer smaller than $N$ and checking $(a,N)=1$. If it is not, we have a factor of $N$ and we can declare $N$ composite and give the factor as proof. Before going into the quantum part, we perform a basic screening to reduce the number of quantum operations, which are the most challenging in terms of technology, and replace them by classical steps. 
 
We need to check $a^{N-1}\equiv 1 \mod N$. Instead of performing this Fermat test directly, we check $a^{\frac{N-1}{2}}$, which should be $\pm 1$ if $a$ passes Fermat's test. Otherwise, $N$ cannot be prime and we return $a$ as a primality witness for the Fermat test. If $a^{\frac{N-1}{2}}\equiv 1 \mod N$, the order of $a$ is, at most, $\frac{N-1}{2}$, which gives no information on whether $N$ is prime or not. Then we start again and choose a new random integer. We only proceed if $a^{\frac{N-1}{2}}\equiv -1 \mod N$. 

At this point, we need to resort to the quantum order finding algorithm. If $\ord(a)=N-1$ the number is prime with certainty and we can stop the procedure and return $a$ as a quantum certificate of primality. 

The classical screening guarantees $\ord(a)| N-1$. This follows from the condition $a^{\frac{N-1}{2}}\equiv -1 \mod N$, which means $a^{N-1}\equiv 1 \mod N$, and from \cite{Bur05}:
\begin{thm}\label{OrderDiv}
Let $a \in \mathbb{Z}_N^*$ have order $\ord(a)$. Then $a^h\equiv 1\mod N$ if and only if $\ord(a)| h$.
\end{thm}
If $\ord(a)\neq N-1$, we cannot tell anything about $N$. We need to start again the search with a new element. 

The average number of iterations before finding an element of order $N-1$ when $N$ is prime is of the order of $\log\log N$. From Theorem \ref{OrderPrimesTh}, we know there are $\varphi(N-1)$ elements of order $N-1$ among the $N-1$ elements of $\mathbb{Z}_N^*$. The probability of finding an element which confirms primality at each iteration is
\begin{equation}
\label{DensitySols}
\frac{\varphi(N-1)}{N-1}> \frac{1}{3\log\log (N-1)}
\end{equation}
for a large enough $N$.
In order to prove this bound we can turn to known estimations, starting from the lower bound \cite{RS62}:
\begin{equation}
\label{LowRosserSchonfeld}
\frac{\varphi(M)}{M}> \frac{1}{e^{\gamma}\log\log M+\frac{2.50637}{\log\log M}}
\end{equation}
where $\gamma\approx 0.57721$ is the Euler-Mascheroni constant and $e^\gamma\approx 1.781$. For $M>e^{e^{\sqrt{\frac{2.50637}{3-1.781}}}} \approx 49.2$ there is a lower bound
\begin{equation}
\frac{\varphi(M)}{M}> \frac{1}{3\log\log M},
\end{equation}
which tells us that for a prime $N$ of $n\geq 6$ bits Equation ($\ref{DensitySols}$) will be valid. With $3\log\log (N-1)$ attempts we have a high probability of finding an element of order $N-1$. While there is some room to improve the estimate, in the general case we cannot give a much tighter bound as it is known that for infinitely many integers
\begin{equation}
\frac{\varphi(M)}{M}< \frac{1}{e^\gamma \log\log M}<\frac{0.562}{\log\log M}
\end{equation}
will hold \cite{Nic83}. We need a number of repetitions logarithmic with the number of bits of the integer under test. The complexity of each iteration is determined by the order finding subroutine. 

Quantum order finding requires $O((\log n )n^3)$ quantum operations, with $O(\log n)$ uses of modular exponentiation. The quantum order finding subroutine of Shor has two main steps: modular exponentiation and a Quantum Fourier Transform. The Quantum Fourier Transform circuit is quadratic in $n$ \cite{Sho97}. Modular exponentiation with the binary method needs $O(n)$ multiplications \cite{Knu97}. There are many quadratic quantum multiplication circuits, for instance \cite{VBE96,BCD96}, which gives a total complexity of $O(n^3)$ for exponentiation. In principle, with fast multiplication using the Sch\"onhage-Strassen algorithm \cite{SS71}, for which there is a quantum circuit \cite{Zal98}, the total complexity for order finding would be $O(\log\log n (\log n)^2 n^2)$. However, the constant factors involved make it only worthwhile in the asymptotic limit for very large $N$ \cite{vMI05}.

The number of operations in the classical part is also dominated by modular exponentiation. Computing the greatest common divisor of two integers up to $n$ bits using Euclid's algorithm has a complexity $O(n^2)$ and there are faster modern methods (see chapter 4 of \cite{BS96}). The total expected complexity of our algorithm is $O((\log n)^2 n^3)$ for the $\log n$ repetitions needed to find an element of order $N-1$ with high probability. For very large $N$ we can use fast multiplication to have an expected number of operations $O(\log\log n (\log n)^3 n^2)$.

The screening in line \ref{EulerTest} of the algorithm identifies $N$ is not prime when $\ord(a)\not\hspace{-0.8ex}| \hspace{0.8ex}N-1$, but it is possible for $N$ to be composite and still give inconclusive results when tested. For instance, Carmichael numbers satisfy $a^{N-1}\equiv 1 \mod N$ for all $a$ such that $(a,N)=1$ and, from Theorem $\ref{OrderDiv}$, $\ord(a)| N-1$ for all $a \in \mathbb{Z}_N^*$. In any case, the order will be smaller than $N-1$ and, after $3\log n$ tested elements, we can say that $N$ is composite with high probability and stop there to avoid entering an infinite loop.

In order to reduce the number of quantum operations, we can introduce a previous classical selection phase that uses the Miller-Rabin test. The elements $a \in \mathbb{Z}_N^*$ for which $a^{\frac{N-1}{2}}\equiv \pm 1 \mod N$ for a composite $N$ are sometimes called Euler liars \footnote{The most usual definition of Euler liars is the $a$ which satisfy $a^{\frac{N-1}{2}}\equiv\left(\frac{a}{N}\right)  \mod N$ for a composite $N$, where $\left(\frac{a}{N}\right)$ is the Jacobi symbol.}. Euler liars are a subgroup of Fermat liars, the $a$ for which a composite $N$ passes the Fermat test. We can impose harder constraints on the integers that survive to the quantum part of the algorithm. For a composite number $N$ with $N-1=2^s d$, with odd $d$, the bases for which $a^d\equiv 1 \mod N$ or $a^{d2^r}\equiv -1 \mod N$ for some $0 \leq r < s$ are called strong liars. For a prime $N$ the condition always holds, but, if $N$ is composite, at most one fourth of the $a \in \mathbb{Z}_N^*$ are strong liars \cite{Rab80}. If we perform a classical Rabin-Miller test on $k$ random bases and do not find a witness for compositeness, the probability of $N$ not being prime is bounded by $4^{-k}$ and we are left with a collection of $k$ elements of order $\ord(a)|N-1$. We can discard the bases with $a^d\equiv 1 \mod N$, which have order $d<N-1$, and use the rest of the elements in the quantum order finding subroutine.

Finally, we can further reduce the number of steps with some insights from the analysis of quantum order finding. The factor $\log n$ which appears in the complexity of quantum order finding is due to the average number of times we have to measure in order to find two divisors of the order from which we can deduce its exact value, $\ord(a)$. However, with some classical processing testing small multiples of the values extracted from each measurement, it is possible to reduce the $\log n$ repetitions to a constant number \cite{Sho97}. 

The algorithm we have proposed offers an alternative quantum primality test which harnesses quantum order finding to give a direct proof an integer is prime by producing an element $a \in \mathbb{Z}_N^*$ with order $N-1$. The quantum part uses the same circuits as Shor's factoring algorithm and it could serve as a previous stage when factoring on a quantum machine. Shor's algorithm requires its input integer $N$ not to be of the form $p^k$ or $2p^k$ for a prime $p$ and an integer $k\geq 1$. For odd inputs, we only need to worry about detecting primes. There are classical efficient methods to detect prime powers $p^k$ for $k\geq 2$ \cite{Ber98}, but we can also use a modified version of our quantum primality test. Theorems \ref{OrderTh} and \ref{OrderPrimesTh} can be generalized to show that an integer $N>1$ has an element of order $\varphi(N)$, called a \emph{primitive root}, only when $N=2,4, p^k$ or $p^{2k}$, in which case there are $\varphi(\varphi(N))$ of them (chapter 4 of \cite{Bur05}). The analysis then is essentially the same we have used in our primality test. For $N=p^k$, $\varphi(N)=p^{k-1}(p-1)$. If we find a primitive root, its order $\ord(a)$ gives $(\ord(a),N)=p^{k-1}\neq 1$, which factors $N$. We can check by repeated division by $p=\frac{N}{(\ord(a),N)}$ that $N$ is a prime power. The number of divisions is polynomial in the number of bits of $N$ and the bound of $\log n$ order finding steps is still valid. The probability of finding a primitive root is $\frac{\varphi(\varphi(N))}{\varphi(N)}$ and $3 \log\log\varphi(N)$ repetitions give a high probability of getting a valid basis. $\varphi(N)\leq N-1$, so our bound for the primes also holds in this situation.

Our algorithm reduces the asymptotic complexity of the Chau-Lo quantum primality test from $O( (\log n) (\log\log n) n^3)$ to $O((\log\log n) (\log n)^3 n^2 )$ at the cost of replacing the classical primality certificate which includes the factors of $N-1$ by a quantum certificate of primality consisting in an element $a$ of order $N-1$, which can be checked on a quantum computer. The test can prove with certainty that a given integer is prime and it can be complemented with the Miller-Rabin test in the initial screening stage to also identify composite numbers with high probability. Our test has a complexity comparable to classical tests for compositeness which can convince us a number is prime with an exponentially small probability of error. Its complexity is essentially quadratic in the asymptotic limit, which is more efficient than classical tests that prove primality with certainty, which are usually restricted to integers of a particular form, or require a number of operations of the order of the sixth power of the number of bits (AKS test).

\section*{Acknowledgements} 
This work has been funded by Spanish Ministerio de Econom\'ia y Competitividad, Project TEC2015-69665-R, MINECO/FEDER, UE and Junta de Castilla y Le\'on VA089U16.
 

\begin{thebibliography}{31}%
\makeatletter
\providecommand \@ifxundefined [1]{%
 \@ifx{#1\undefined}
}%
\providecommand \@ifnum [1]{%
 \ifnum #1\expandafter \@firstoftwo
 \else \expandafter \@secondoftwo
 \fi
}%
\providecommand \@ifx [1]{%
 \ifx #1\expandafter \@firstoftwo
 \else \expandafter \@secondoftwo
 \fi
}%
\providecommand \natexlab [1]{#1}%
\providecommand \enquote  [1]{``#1''}%
\providecommand \bibnamefont  [1]{#1}%
\providecommand \bibfnamefont [1]{#1}%
\providecommand \citenamefont [1]{#1}%
\providecommand \href@noop [0]{\@secondoftwo}%
\providecommand \href [0]{\begingroup \@sanitize@url \@href}%
\providecommand \@href[1]{\@@startlink{#1}\@@href}%
\providecommand \@@href[1]{\endgroup#1\@@endlink}%
\providecommand \@sanitize@url [0]{\catcode `\\12\catcode `\$12\catcode
  `\&12\catcode `\#12\catcode `\^12\catcode `\_12\catcode `\%12\relax}%
\providecommand \@@startlink[1]{}%
\providecommand \@@endlink[0]{}%
\providecommand \url  [0]{\begingroup\@sanitize@url \@url }%
\providecommand \@url [1]{\endgroup\@href {#1}{\urlprefix }}%
\providecommand \urlprefix  [0]{URL }%
\providecommand \Eprint [0]{\href }%
\providecommand \doibase [0]{http://dx.doi.org/}%
\providecommand \selectlanguage [0]{\@gobble}%
\providecommand \bibinfo  [0]{\@secondoftwo}%
\providecommand \bibfield  [0]{\@secondoftwo}%
\providecommand \translation [1]{[#1]}%
\providecommand \BibitemOpen [0]{}%
\providecommand \bibitemStop [0]{}%
\providecommand \bibitemNoStop [0]{.\EOS\space}%
\providecommand \EOS [0]{\spacefactor3000\relax}%
\providecommand \BibitemShut  [1]{\csname bibitem#1\endcsname}%
\let\auto@bib@innerbib\@empty
\bibitem [{\citenamefont {Menezes}\ \emph {et~al.}(1996)\citenamefont
  {Menezes}, \citenamefont {Vanstone},\ and\ \citenamefont {Oorschot}}]{MVS96}%
  \BibitemOpen
  \bibfield  {author} {\bibinfo {author} {\bibfnamefont {A.~J.}\
  \bibnamefont {Menezes}}, \bibinfo {author} {\bibfnamefont {S.~A.}\
  \bibnamefont {Vanstone}}, \ and\ \bibinfo {author} {\bibfnamefont {P.
  C.}\ \bibnamefont {van Oorschot}},\ }\href@noop {} {\emph {\bibinfo {title}
  {Handbook of Applied Cryptography}}},\ \bibinfo {edition} {1st}\ ed.\
  (\bibinfo  {publisher} {CRC Press, Inc.},\ \bibinfo {address} {Boca Raton,
  FL, USA},\ \bibinfo {year} {1996})\BibitemShut {NoStop}%
\bibitem [{\citenamefont {Burton}(2005)}]{Bur05}%
  \BibitemOpen
  \bibfield  {author} {\bibinfo {author} {\bibfnamefont {D.~M.}\
  \bibnamefont {Burton}},\ }\href@noop {} {\emph {\bibinfo {title} {{Elementary
  Number Theory}}}},\ \bibinfo {edition} {6th}\ ed.\ (\bibinfo  {publisher}
  {McGraw-Hill Higher Education},\ \bibinfo {year} {2005})\BibitemShut
  {NoStop}%
\bibitem [{\citenamefont {Crandall}\ and\ \citenamefont
  {Pomerance}(2005)}]{CP05}%
  \BibitemOpen
  \bibfield  {author} {\bibinfo {author} {\bibfnamefont {R.}\ \bibnamefont
  {Crandall}}\ and\ \bibinfo {author} {\bibfnamefont {C.}\ \bibnamefont
  {Pomerance}},\ }\href {\doibase 10.1007/0-387-28979-8} {\emph {\bibinfo
  {title} {Prime Numbers: A Computational Perspective}}}\ (\bibinfo
  {publisher} {Springer New York},\ \bibinfo {address} {New York, NY},\
  \bibinfo {year} {2005})\BibitemShut {NoStop}%
\bibitem [{\citenamefont {Carmichael}(1912)}]{Car12}%
  \BibitemOpen
  \bibfield  {author} {\bibinfo {author} {\bibfnamefont {R.~D.}\ \bibnamefont
  {Carmichael}},\ }\bibfield  {title} {\enquote {\bibinfo {title} {On composite
  numbers {P} which satisfy the {F}ermat congruence $a^{P-1} \equiv 1 \mod
  P$},}\ }\href{\doibase 10.2307/2972687} {\bibfield  {journal} {\bibinfo  {journal} {The
  American Mathematical Monthly}\ }\textbf {\bibinfo {volume} {19}},\ \bibinfo
  {pages} {22--27} (\bibinfo {year} {1912})}\BibitemShut {NoStop}%
\bibitem [{\citenamefont {Alford}\ \emph {et~al.}(1994)\citenamefont {Alford},
  \citenamefont {Granville},\ and\ \citenamefont {Pomerance}}]{AGP94}%
  \BibitemOpen
  \bibfield  {author} {\bibinfo {author} {\bibfnamefont {W.~R.}\ \bibnamefont
  {Alford}}, \bibinfo {author} {\bibfnamefont {A.}\ \bibnamefont
  {Granville}}, \ and\ \bibinfo {author} {\bibfnamefont {C.}\ \bibnamefont
  {Pomerance}},\ }\bibfield  {title} {\enquote {\bibinfo {title} {There are
  infinitely many {C}armichael numbers},}\ }\href{\doibase 10.2307/2118576} {\bibfield  {journal}
  {\bibinfo  {journal} {Annals of Mathematics}\ }\textbf {\bibinfo {volume}
  {139}},\ \bibinfo {pages} {703--722} (\bibinfo {year} {1994})}\BibitemShut
  {NoStop}%
\bibitem [{\citenamefont {Solovay}\ and\ \citenamefont
  {Strassen}(1977)}]{SS77}%
  \BibitemOpen
  \bibfield  {author} {\bibinfo {author} {\bibfnamefont {R.}~\bibnamefont
  {Solovay}}\ and\ \bibinfo {author} {\bibfnamefont {V.}~\bibnamefont
  {Strassen}},\ }\bibfield  {title} {\enquote {\bibinfo {title} {A fast
  {M}onte-{C}arlo test for primality},}\ }\href {\doibase 10.1137/0206006}
  {\bibfield  {journal} {\bibinfo  {journal} {SIAM Journal on Computing}\
  }\textbf {\bibinfo {volume} {6}},\ \bibinfo {pages} {84--85} (\bibinfo {year}
  {1977})}\BibitemShut {NoStop}%
\bibitem [{\citenamefont {Miller}(1975)}]{Mil75}%
  \BibitemOpen
  \bibfield  {author} {\bibinfo {author} {\bibfnamefont {G.~L.}\ \bibnamefont
  {Miller}},\ }\bibfield  {title} {\enquote{\bibinfo {title} {Riemann's
  hypothesis and tests for primality},}\ }in\ \href {\doibase
  10.1145/800116.803773} {\emph {\bibinfo {booktitle} {Proceedings of the
  Seventh Annual ACM Symposium on Theory of Computing}}},\ \bibinfo {series and
  number} {STOC '75}\ (\bibinfo  {publisher} {ACM},\ \bibinfo {address} {New
  York, NY, USA},\ \bibinfo {year} {1975})\ pp.\ \bibinfo {pages}
  {234--239}\BibitemShut {NoStop}%
\bibitem [{\citenamefont {Rabin}(1980)}]{Rab80}%
  \BibitemOpen
  \bibfield  {author} {\bibinfo {author} {\bibfnamefont {M.~O.}\
  \bibnamefont {Rabin}},\ }\bibfield  {title} {\enquote {\bibinfo {title}
  {Probabilistic algorithm for testing primality},}\ }\href {\doibase
  https://doi.org/10.1016/0022-314X(80)90084-0} {\bibfield  {journal} {\bibinfo
   {journal} {Journal of Number Theory}\ }\textbf {\bibinfo {volume} {12}},\
  \bibinfo {pages} {128 -- 138} (\bibinfo {year} {1980})}\BibitemShut {NoStop}%
\bibitem [{\citenamefont {Dixon}(1984)}]{Dix84}%
  \BibitemOpen
  \bibfield  {author} {\bibinfo {author} {\bibfnamefont {J.~D.}\ \bibnamefont
  {Dixon}},\ }\bibfield  {title} {\enquote {\bibinfo {title} {Factorization and
  primality tests},}\ }\href{\doibase 10.2307/2322136} {\bibfield  {journal} {\bibinfo  {journal}
  {The American Mathematical Monthly}\ }\textbf {\bibinfo {volume} {91}},\
  \bibinfo {pages} {333--352} (\bibinfo {year} {1984})}\BibitemShut {NoStop}%
\bibitem [{\citenamefont {Agrawal}\ \emph {et~al.}(2004)\citenamefont
  {Agrawal}, \citenamefont {Kayal},\ and\ \citenamefont {Saxena}}]{AKS04}%
  \BibitemOpen
  \bibfield  {author} {\bibinfo {author} {\bibfnamefont {M.}\
  \bibnamefont {Agrawal}}, \bibinfo {author} {\bibfnamefont {N.}\
  \bibnamefont {Kayal}}, \ and\ \bibinfo {author} {\bibfnamefont {N.}\
  \bibnamefont {Saxena}},\ }\bibfield  {title} {\enquote {\bibinfo {title}
  {{PRIMES} is in {P}},}\ }\href {http://www.jstor.org/stable/3597229}
  {\bibfield  {journal} {\bibinfo  {journal} {Annals of Mathematics}\ }\textbf
  {\bibinfo {volume} {160}},\ \bibinfo {pages} {781--793} (\bibinfo {year}
  {2004})}\BibitemShut {NoStop}%
\bibitem [{\citenamefont {Lenstra}\ and\ \citenamefont
  {Pomerance}(2003)}]{LP03}%
  \BibitemOpen
  \bibfield  {author} {\bibinfo {author} {\bibfnamefont {H.~W.}\ \bibnamefont
  {Lenstra}}\ and\ \bibinfo {author} {\bibfnamefont {C.}~\bibnamefont
  {Pomerance}},\ }\bibfield  {title} {\enquote {\bibinfo {title} {Primality
  testing with {G}aussian periods},}\ }\href@noop {} {\bibfield  {journal}
  {\bibinfo  {journal} {preprint:
  \href{https://www.math.dartmouth.edu/~carlp/aks240817.pdf}{https://www.math.dartmouth.edu/~carlp/aks240817.pdf}}\ }
  (\bibinfo {year} {2003})}\BibitemShut {NoStop}%
\bibitem [{\citenamefont {Lucas}(1878)}]{Luc78}%
  \BibitemOpen
  \bibfield  {author} {\bibinfo {author} {\bibfnamefont {E.}\ \bibnamefont
  {Lucas}},\ }\bibfield  {title} {\enquote {\bibinfo {title} {Th\'eorie des
  fonctions num\'eriques simplement p\'eriodiques},}\ }\href{\doibase 10.2307/2369308} {\bibfield
   {journal} {\bibinfo  {journal} {American Journal of Mathematics}\ }\textbf
  {\bibinfo {volume} {1}},\ \bibinfo {pages} {289--321} (\bibinfo {year}
  {1878})}\BibitemShut {NoStop}%
\bibitem [{\citenamefont {Lehmer}(1927)}]{Leh27}%
  \BibitemOpen
  \bibfield  {author} {\bibinfo {author} {\bibfnamefont {D.~H.}\ \bibnamefont
  {Lehmer}},\ }\bibfield  {title} {\enquote {\bibinfo {title} {Tests for
  primality by the converse of {F}ermat's theorem},}\ }\href{\doibase 10.1090/S0002-9904-1927-04368-3} {\bibfield
   {journal} {\bibinfo  {journal} {Bulletin of the American Mathematical
  Society}\ }\textbf {\bibinfo {volume} {33}},\ \bibinfo {pages} {327--340}
  (\bibinfo {year} {1927})}\BibitemShut {NoStop}%
\bibitem [{\citenamefont {Brillhart}\ \emph {et~al.}(1975)\citenamefont
  {Brillhart}, \citenamefont {Lehmer},\ and\ \citenamefont
  {Selfridge}}]{BLS75}%
  \BibitemOpen
  \bibfield  {author} {\bibinfo {author} {\bibfnamefont {John}\ \bibnamefont
  {Brillhart}}, \bibinfo {author} {\bibfnamefont {D.~H.}\ \bibnamefont
  {Lehmer}}, \ and\ \bibinfo {author} {\bibfnamefont {J.~L.}\ \bibnamefont
  {Selfridge}},\ }\bibfield  {title} {\enquote {\bibinfo {title} {New primality
  criteria and factorizations of $2^m \pm 1$},}\ }\href{\doibase 10.1090/S0025-5718-1975-0384673-1} {\bibfield
  {journal} {\bibinfo  {journal} {Mathematics of Computation}\ }\textbf
  {\bibinfo {volume} {29}},\ \bibinfo {pages} {620--647} (\bibinfo {year}
  {1975})}\BibitemShut {NoStop}%
\bibitem [{\citenamefont {work}(Ongoing from 1996)}]{GIMPS}%
  \BibitemOpen
  \bibfield  {author} {\bibinfo {author} {\bibfnamefont {Collective}\
  \bibnamefont {work}},\ }\href@noop {} {\enquote {\bibinfo {title} {{The Great
  Internet Mersenne Prime Search (GIMPS)}},}\ }{\bibinfo  {journal} {online:
  \href{https://www.mersenne.org/}{https://www.mersenne.org/}}\ } (\bibinfo {year} {Ongoing from
  1996})\BibitemShut {NoStop}%
\bibitem [{\citenamefont {Pratt}(1975)}]{Pra75}%
  \BibitemOpen
  \bibfield  {author} {\bibinfo {author} {\bibfnamefont {V.~R.}\
  \bibnamefont {Pratt}},\ }\bibfield  {title} {\enquote {\bibinfo {title}
  {Every prime has a succinct certificate},}\ }\href {\doibase 10.1137/0204018}
  {\bibfield  {journal} {\bibinfo  {journal} {SIAM Journal on Computing}\
  }\textbf {\bibinfo {volume} {4}},\ \bibinfo {pages} {214--220} (\bibinfo
  {year} {1975})}\BibitemShut {NoStop}%
\bibitem [{\citenamefont {Pomerance}(1987)}]{Pom87}%
  \BibitemOpen
  \bibfield  {author} {\bibinfo {author} {\bibfnamefont {C.}\ \bibnamefont
  {Pomerance}},\ }\bibfield  {title} {\enquote {\bibinfo {title} {Very short
  primality proofs},}\ }\href{\doibase 10.1090/S0025-5718-1987-0866117-4 } {\bibfield  {journal} {\bibinfo
  {journal} {Mathematics of Computation}\ }\textbf {\bibinfo {volume} {48}},\
  \bibinfo {pages} {315--322} (\bibinfo {year} {1987})}\BibitemShut {NoStop}%
\bibitem [{\citenamefont {Shor}(1997)}]{Sho97}%
  \BibitemOpen
  \bibfield  {author} {\bibinfo {author} {\bibfnamefont {P.~W.}\
  \bibnamefont {Shor}},\ }\bibfield  {title} {\enquote {\bibinfo {title}
  {Polynomial-time algorithms for prime factorization and discrete logarithms
  on a quantum computer},}\ }\href{\doibase 10.1137/S0097539795293172}{\bibfield  {journal} {\bibinfo
  {journal} {\mbox{SIAM Journal on Computing}}\ }\textbf {\bibinfo {volume}
  {26}},\ \bibinfo {pages} {1484} (\bibinfo {year} {1997})}\BibitemShut
  {NoStop}%
\bibitem [{\citenamefont {Chau}\ and\ \citenamefont {Lo}(1997)}]{CL97}%
  \BibitemOpen
  \bibfield  {author} {\bibinfo {author} {\bibfnamefont {H.~F.}\ \bibnamefont
  {Chau}}\ and\ \bibinfo {author} {\bibfnamefont {H.-K.}\ \bibnamefont {Lo}},\
  }\bibfield  {title} {\enquote {\bibinfo {title} {Primality test via quantum
  factorization},}\ }\href {\doibase 10.1142/S0129183197000138} {\bibfield
  {journal} {\bibinfo  {journal} {International Journal of Modern Physics C}\
  }\textbf {\bibinfo {volume} {08}},\ \bibinfo {pages} {131--138} (\bibinfo
  {year} {1997})}\BibitemShut {NoStop}%
\bibitem [{\citenamefont {Carlini}\ and\ \citenamefont {Hosoya}(2000)}]{CH00}%
  \BibitemOpen
  \bibfield  {author} {\bibinfo {author} {\bibfnamefont {A.}~\bibnamefont
  {Carlini}}\ and\ \bibinfo {author} {\bibfnamefont {A.}~\bibnamefont
  {Hosoya}},\ }\bibfield  {title} {\enquote {\bibinfo {title} {Quantum
  probabilistic subroutines and problems in number theory},}\ }\href {\doibase
  10.1103/PhysRevA.62.032312} {\bibfield  {journal} {\bibinfo  {journal}
  {Physical Review A}\ }\textbf {\bibinfo {volume} {62}},\ \bibinfo {pages}
  {032312} (\bibinfo {year} {2000})}\BibitemShut {NoStop}%
\bibitem [{\citenamefont {Rosser}\ and\ \citenamefont
  {Schoenfeld}(1962)}]{RS62}%
  \BibitemOpen
  \bibfield  {author} {\bibinfo {author} {\bibfnamefont {J.~B.}\
  \bibnamefont {Rosser}}\ and\ \bibinfo {author} {\bibfnamefont {L.}\
  \bibnamefont {Schoenfeld}},\ }\bibfield  {title} {\enquote {\bibinfo {title}
  {Approximate formulas for some functions of prime numbers},}\ }\href@noop {}
  \href{https://projecteuclid.org/euclid.ijm/1255631807}{\bibfield  {journal} {\bibinfo  {journal} {Illinois Journal of Mathematics}\
  }\textbf {\bibinfo {volume} {6}},\ \bibinfo {pages} {64--94} (\bibinfo {year}
  {1962})}\BibitemShut {NoStop}%
\bibitem [{\citenamefont {Nicolas}(1983)}]{Nic83}%
  \BibitemOpen
  \bibfield  {author} {\bibinfo {author} {\bibfnamefont {J.-L.}\
  \bibnamefont {Nicolas}},\ }\bibfield  {title} {\enquote {\bibinfo {title}
  {Petites valeurs de la fonction d'{E}uler},}\ }\href {\doibase
  https://doi.org/10.1016/0022-314X(83)90055-0} {\bibfield  {journal} {\bibinfo
   {journal} {Journal of Number Theory}\ }\textbf {\bibinfo {volume} {17}},\
  \bibinfo {pages} {375 -- 388} (\bibinfo {year} {1983})}\BibitemShut {NoStop}%
\bibitem [{\citenamefont {Knuth}(1997)}]{Knu97}%
  \BibitemOpen
  \bibfield  {author} {\bibinfo {author} {\bibfnamefont {D.~E.}\
  \bibnamefont {Knuth}},\ }\href@noop {} {\emph {\bibinfo {title} {{The Art of
  Computer Programming, Volume 2 (3rd Ed.): Seminumerical Algorithms}}}}\
  (\bibinfo  {publisher} {Addison-Wesley Longman Publishing Co., Inc.},\
  \bibinfo {address} {Boston, MA, USA},\ \bibinfo {year} {1997})\BibitemShut
  {NoStop}%
\bibitem [{\citenamefont {Vedral}\ \emph {et~al.}(1996)\citenamefont {Vedral},
  \citenamefont {Barenco},\ and\ \citenamefont {Ekert}}]{VBE96}%
  \BibitemOpen
  \bibfield  {author} {\bibinfo {author} {\bibfnamefont {V.}\ \bibnamefont
  {Vedral}}, \bibinfo {author} {\bibfnamefont {A.}\ \bibnamefont
  {Barenco}}, \ and\ \bibinfo {author} {\bibfnamefont {A.}\ \bibnamefont
  {Ekert}},\ }\bibfield  {title} {\enquote {\bibinfo {title} {Quantum networks
  for elementary arithmetic operations},}\ }\href {\doibase
  10.1103/PhysRevA.54.147} {\bibfield  {journal} {\bibinfo  {journal} {Physical
  Review A}\ }\textbf {\bibinfo {volume} {54}},\ \bibinfo {pages} {147--153}
  (\bibinfo {year} {1996})}\BibitemShut {NoStop}%
\bibitem [{\citenamefont {Beckman}\ \emph {et~al.}(1996)\citenamefont
  {Beckman}, \citenamefont {Chari}, \citenamefont {Devabhaktuni},\ and\
  \citenamefont {Preskill}}]{BCD96}%
  \BibitemOpen
  \bibfield  {author} {\bibinfo {author} {\bibfnamefont {D.}\ \bibnamefont
  {Beckman}}, \bibinfo {author} {\bibfnamefont {A.~N.}\ \bibnamefont
  {Chari}}, \bibinfo {author} {\bibfnamefont {S.}\ \bibnamefont
  {Devabhaktuni}}, \ and\ \bibinfo {author} {\bibfnamefont {J.}\ \bibnamefont
  {Preskill}},\ }\bibfield  {title} {\enquote {\bibinfo {title} {Efficient
  networks for quantum factoring},}\ }\href {\doibase 10.1103/PhysRevA.54.1034}
  {\bibfield  {journal} {\bibinfo  {journal} {Phys. Rev. A}\ }\textbf {\bibinfo
  {volume} {54}},\ \bibinfo {pages} {1034--1063} (\bibinfo {year}
  {1996})}\BibitemShut {NoStop}%
\bibitem [{\citenamefont {Sch{\"o}nhage}\ and\ \citenamefont
  {Strassen}(1971)}]{SS71}%
  \BibitemOpen
  \bibfield  {author} {\bibinfo {author} {\bibfnamefont {A.}~\bibnamefont
  {Sch{\"o}nhage}}\ and\ \bibinfo {author} {\bibfnamefont {V.}~\bibnamefont
  {Strassen}},\ }\bibfield  {title} {\enquote {\bibinfo {title} {Schnelle
  {M}ultiplikation gro{\ss}er {Z}ahlen},}\ }\href {\doibase 10.1007/BF02242355}
  {\bibfield  {journal} {\bibinfo  {journal} {Computing}\ }\textbf {\bibinfo
  {volume} {7}},\ \bibinfo {pages} {281--292} (\bibinfo {year}
  {1971})}\BibitemShut {NoStop}%
\bibitem [{\citenamefont {Zalka}(1998)}]{Zal98}%
  \BibitemOpen
  \bibfield  {author} {\bibinfo {author} {\bibfnamefont {C.}\
  \bibnamefont {Zalka}},\ }\bibfield  {title} {\enquote {\bibinfo {title}
  {{Fast versions of Shor's quantum factoring algorithm}},}\ }\href@noop {}
  {\bibfield  {journal} {\bibinfo  {journal} preprint \href {https://arxiv.org/abs/quant-ph/9806084}}{quant-ph/9806084\ }
  (\bibinfo {year} {1998})}\BibitemShut {NoStop}%
\bibitem [{\citenamefont {Van~Meter}\ and\ \citenamefont {Itoh}(2005)}]{vMI05}%
  \BibitemOpen
  \bibfield  {author} {\bibinfo {author} {\bibfnamefont {R.}\ \bibnamefont
  {Van~Meter}}\ and\ \bibinfo {author} {\bibfnamefont {K.~M.}\ \bibnamefont
  {Itoh}},\ }\bibfield  {title} {\enquote {\bibinfo {title} {Fast quantum
  modular exponentiation},}\ }\href {\doibase 10.1103/PhysRevA.71.052320}
  {\bibfield  {journal} {\bibinfo  {journal} {Physical Review A}\ }\textbf
  {\bibinfo {volume} {71}},\ \bibinfo {pages} {052320} (\bibinfo {year}
  {2005})}\BibitemShut {NoStop}%
\bibitem [{\citenamefont {Bach}\ and\ \citenamefont {Shallit}(1996)}]{BS96}%
  \BibitemOpen
  \bibfield  {author} {\bibinfo {author} {\bibfnamefont {E.}\ \bibnamefont
  {Bach}}\ and\ \bibinfo {author} {\bibfnamefont {J.}\ \bibnamefont
  {Shallit}},\ }\href@noop {} {\emph {\bibinfo {title} {Algorithmic Number
  Theory; Volume {I}: Efficient Algorithms}}}\ (\bibinfo  {publisher} {The MIT
  Press},\ \bibinfo {year} {1996})\BibitemShut {NoStop}%
\bibitem [{Note1()}]{Note1}%
  \BibitemOpen
  \bibinfo {note} {The most usual definition of Euler liars is the $a$ which
  satisfy $a^{\protect \frac {N-1}{2}}\equiv \left (\protect \frac {a}{N}\right
  ) \penalty \z@ \mkern 12mu{\mathgroup \symoperators mod}\protect \tmspace
  +\thinmuskip {.1667em}\protect \tmspace +\thinmuskip {.1667em}N$ for a
  composite $N$, where $\left (\protect \frac {a}{N}\right )$ is the Jacobi
  symbol.}\BibitemShut {Stop}%
\bibitem [{\citenamefont {Bernstein}(1998)}]{Ber98}%
  \BibitemOpen
  \bibfield  {author} {\bibinfo {author} {\bibfnamefont {D.~J.}\
  \bibnamefont {Bernstein}},\ }\bibfield  {title} {\enquote {\bibinfo {title}
  {Detecting perfect powers in essentially linear time},}\ }\href@noop {}
  \href {\doibase 10.1090/S0025-5718-98-00952-1}{\bibfield  {journal} {\bibinfo  {journal} {Mathematics of Computation}\
  }\textbf {\bibinfo {volume} {67}},\ \bibinfo {pages} {1253--1283} (\bibinfo
  {year} {1998})}\BibitemShut {NoStop}%
\end{thebibliography}
\newcommand{\noopsort}[1]{} \newcommand{\printfirst}[2]{#1}
  \newcommand{\singleletter}[1]{#1} \newcommand{\switchargs}[2]{#2#1}
\end{document}